\documentclass[journal=nalefd,manuscript=letter]{achemso}

\usepackage{bm}
\usepackage{graphicx}
\usepackage{setspace}
\usepackage[caption=false]{subfig}
\usepackage{wrapfig}
\usepackage{epstopdf}
\usepackage{xcolor}
\usepackage{siunitx}
\usepackage{braket}
\usepackage{tabularx}
\usepackage{blindtext}
\usepackage{amsmath}
\usepackage{float}
\usepackage{tabularx}
\usepackage{multirow}
\usepackage{gensymb} 
\DeclareMathOperator\erfc{erfc} 
\DeclareMathOperator\erf{erf} 
\usepackage{hyperref}
\usepackage{color}
\usepackage[section]{placeins}

\usepackage{xr}
\def\bem#1{\begin{mathletters}\label{#1}}
\def\eml{\end{mathletters}}

\def\4#1{{\boldsymbol{#1}}}
\def\8#1{{\widetilde{#1}}}

\author{Hanna T. Fridman}
\altaffiliation{These authors contributed equally.}
\affiliation{Institute of Applied Physics, The Hebrew University of Jerusalem, Jerusalem 91904, Israel}

\author{Rotem Malkinson}
\altaffiliation{These authors contributed equally.}
\affiliation{Institute of Applied Physics, The Hebrew University of Jerusalem, Jerusalem 91904, Israel}

\author{Amir Hen}
\affiliation{Institute of Applied Physics, The Hebrew University of Jerusalem, Jerusalem 91904, Israel}

 \author{Shira Yochelis}
 \affiliation{Institute of Applied Physics, The Hebrew University of Jerusalem, Jerusalem 91904, Israel}

\author{Yossi Paltiel}
\affiliation{Institute of Applied Physics, The Hebrew University of Jerusalem, Jerusalem 91904, Israel}
\email{paltiel@mail.huji.ac.il}

\author{Nir Bar-gill}
\affiliation{Institute of Applied Physics, The Hebrew University of Jerusalem, Jerusalem 91904, Israel}
\alsoaffiliation{The Racah Institute of Physics, The Hebrew University of Jerusalem, Jerusalem 91904, Israel}
\email{nir.bar-gill@mail.huji.ac.il}

\title{
Signatures of Spin Coherence in Chiral Coupled Quantum Dots}

\begin{document}


\begin{abstract}
Chiral-induced spin selectivity (CISS) enables spin selectivity of charge carriers in chiral molecular systems without magnetic materials. While spin selectivity has been widely investigated, its quantum coherence has not yet been explored. Here, we investigate spin-dependent photoluminescence (PL) dynamics in multilayer quantum-dot (QD) assemblies coupled by chiral linkers. Using circularly polarized excitation in the presence of an external magnetic field, we observe a pronounced modulation of the PL lifetime that depends on the magnetic field magnitude and geometry. The lifetime difference between left- and right-circularly polarized excitations exhibits a field-angle dependence, consistent with spin precession driven by the transverse magnetic-field component relative to the chiral axis. A model incorporating coupled spin precession and decay processes reproduces the experimental trends. These results establish chiral QD assemblies as a room-temperature platform for probing quantum coherent manifestations of the CISS effect, with implications for spintronic and quantum technologies.
\end{abstract}

\section{Introduction}
Coherent spin manipulation at room temperature is a central goal in the development of scalable quantum technologies. Most existing platforms require cryogenic operation to preserve quantum coherence, which imposes substantial technical and practical limitations. Materials or hybrid structures capable of supporting and measuring coherent spin behavior under ambient conditions could therefore open new pathways for quantum sensing, information processing, and next-generation spintronics\cite{doi:10.1126/science.ads0512,lin_room-temperature_2023,liu_coherent_2019,pirro_advances_2021}.

Chiral materials exist as enantiomers - each enantiomer lacks mirror symmetry, and together they form non-super-imposable mirror-image counterparts. Over the last two decades, extensive research on the chiral-induced spin selectivity (CISS) effect has shown that chiral media strongly distinguish electron spin orientations \cite{bloom_chiral_2024}. Spin selectivity has been demonstrated through asymmetric photoelectron emission from chiral layers \cite{gohler_b_spin_2011,mollers_chirality_2022}, spin-selective electron transfer, and spin or charge-polarized transport through chiral structures \cite{https://doi.org/10.1002/smll.201804557,
naaman_chiral_2020,
annurev:/content/journals/10.1146/annurev-physchem-040214-121554,
xie_spin_2011,
al-bustami_atomic_2022,
bloom_chirality_2017,
aiello_chirality-based_2022}
. These observations highlight the importance of chiral materials in the study of quantum matter. Their strong spin selectivity may support quantum coherent spin transport, suggesting that the study of coherence associated with CISS is essential for both the fundamental understanding of the effect and for potential quantum applications.
Quantum dots (QDs) are widely used in quantum research due to their discrete energy levels and chemical tunability, which allow precise control of their optical and electronic properties\cite{klimov_victor_i_nanocrystal_nodate}. QDs are also considered promising candidates for scalable quantum architectures due to their long coherence times and compatibility with semiconductor fabrication techniques \cite{loss1998quantum,michler2009single,klimov_victor_i_nanocrystal_nodate,press2008complete,warburton2013single,zwanenburg2013silicon,lodahl2015interfacing}. 

Previous work \cite{fridman_spin-exciton_2019} has shown that a chiral medium can mediate coupling between QDs in an asymmetric configuration, resulting in preferential spin delocalization through the chiral linkers. Additional studies \cite{fridman_ultrafast_2023} have reported short-time coherent delocalization in chiral-coupled QDs. Together, these findings indicate that chiral molecules can play a significant role in supporting coherent spin behavior in nanoscale systems.

\begin{figure}[tbh]
    \includegraphics[width = 0.9 \linewidth]{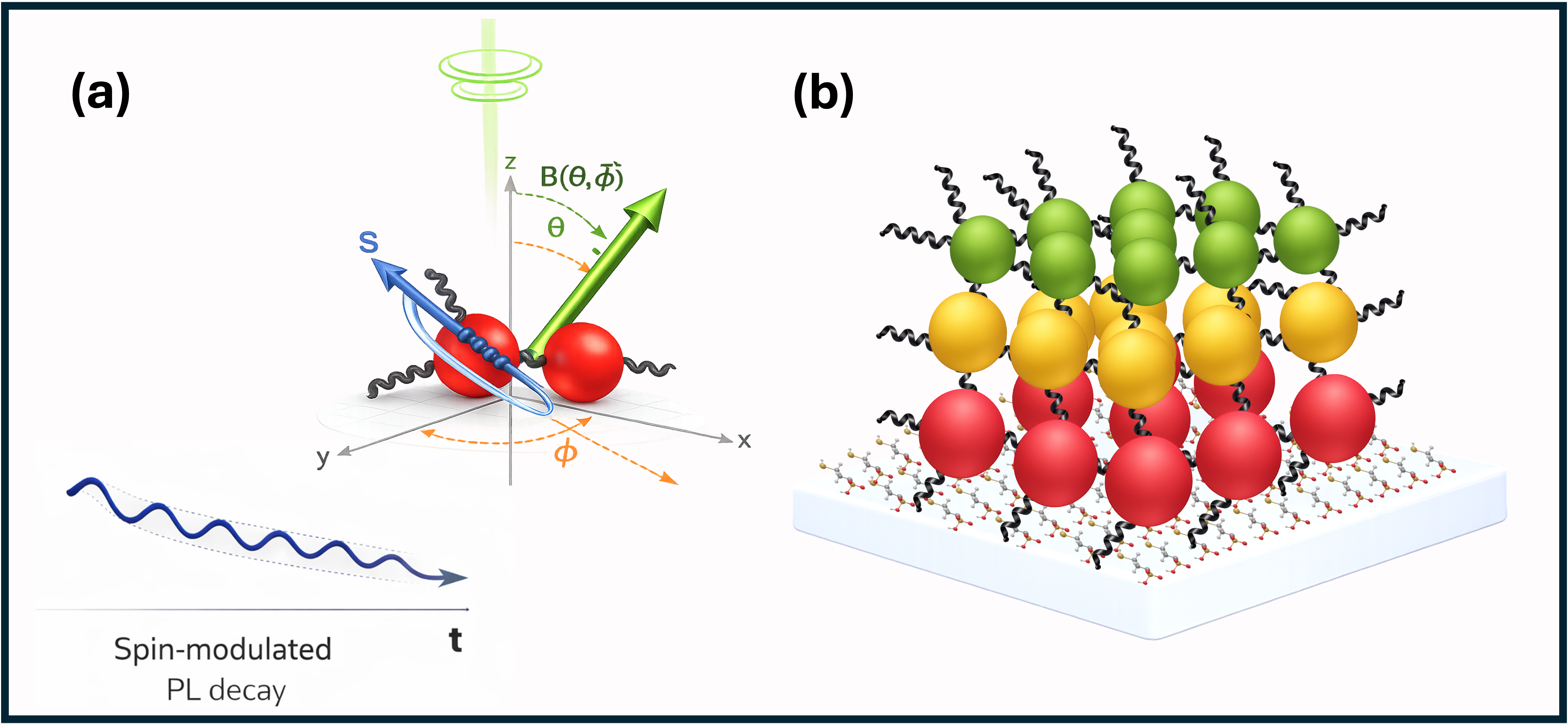}
    \caption{
(a) Schematic illustration of spin-modulated photoluminescence (PL) decay in a chiral–quantum-dot (QD) system, where a transverse magnetic field (B) induces coherent spin precession along the chiral molecular axis, resulting in an oscillatory PL lifetime. 
(b) Schematic of the chiral–QD sample structure, consisting of three QD sizes (3.3, 4.2, and 4.9~nm) with absorption peaks at 520, 560, and 580~nm, respectively, interconnected by left- or right-handed (L/D) chiral molecular linkers.
}
    \label{fig:Illustration_and_QD_chiral_sample}
\end{figure}

In this work, an asymmetric multilayer structure of QDs connected by chiral linkers is used to investigate coherent dynamics due to the CISS effect. The structural asymmetry establishes a directional spin-selective coupling mechanism that gives rise to delocalized spin transport across the QDs. This asymmetric delocalization appears experimentally as a change in the photoluminescence (PL) lifetime for each specific spin orientation \cite{bezen_chiral_2018,fridman_spin-exciton_2019}.
To probe the coherent nature of this effect, we applied a magnetic field to introduce a component perpendicular to the chiral axis. This transverse field induces spin precession in the QD, periodically aligning and misaligning the electron spin with the spin-selective coupling direction (chiral axis)\cite{press2008complete}. As a result, the PL lifetime exhibits an oscillatory dependence on the magnetic-field strength, directly reflecting spin coherence in the chiral–QD system and providing experimental access to the coherent regime of the CISS effect. This behavior is schematically illustrated in Fig.~\ref{fig:Illustration_and_QD_chiral_sample}a.
Importantly, we show that spin-state information can be assessed through photoluminescence lifetime modulation, highlighting chiral QD assemblies as a promising room-temperature platform for probing coherent aspects of the CISS effect, with implications for spintronic and quantum technologies.

\section{Methods}
The chiral structure was constructed using CdSe quantum dots of different sizes and D- or L- $\alpha$-helix polyalanine molecules ([H]-C(AAAAK)7-[OH]). In our structure, the QDs act as artificial atoms \cite{warburton2013single} while the polyalanine molecules serve as the chiral medium \cite{naaman2020acc,fridman_spin-exciton_2019}. By wet chemistry, the QD-chiral-molecule heterostructure was self-assembled layer by layer to produce a multilayer chiral system \cite{decher1997layer}. This results in a network of QDs coupled by the chiral linkers.

Spin selective directed dynamics require breaking of symmetry in the system \cite{fridman_spin-exciton_2019, bezen_chiral_2018}. Therefore, the multilayered QD-chiral-molecule heterostructures were constructed with progressively larger QDs (3.3 nm , 4.2 nm, and 4.9 nm), with absorption peaks at 520, 560, and 580 nm, respectively. The layers were grown in decreasing order of QD size, from large to small, as shown in Fig.~\ref{fig:Illustration_and_QD_chiral_sample}b.
Photoluminescence (PL) lifetimes were measured with a home-built confocal microscope using the time-correlated single-photon counting (TCSPC) technique. This technique constructs a histogram of the arrival times of emitted single photons from a sample excited by a pulsed laser \cite{becker2005tcspc,lakowicz2006principles}. Samples were excited using a 532 nm pulsed laser (OneFive KATANA-05, 50 ps pulse) set to an average power of $\sim$ 250 nW. The laser was circularly polarized to either right or left-handed circular polarization using $\lambda/2$ and $\lambda/4$ waveplates mounted on motorized rotational stages for reproducibility and polarization accuracy. During the measurements, a magnetic field was applied at different angles and strengths, using a static magnet mounted on piezo stages. Before the sample was measured, the magnetic field was characterized and calibrated using Nitrogen-Vacancy (NV) centers in diamond, allowing for precise determination of the exact magnetic field on the sample. In order to reduce background fluorescence noise, the laser was filtered with a 533 nm notch filter along with a 600 nm longpass filter to ensure readout only of the PL signal from the large-size QDs in the chiral sample. A Schematic illustration of the system is presented in Fig.~\ref*{fig:System_schematic_illustration} of the Supporting Information. 

\section{Results}
PL lifetime measurements were performed under magnetic fields ranging between $B = [60,1020]$ Gauss. For all measurements the magnetic field was aligned at a polar angle $\theta = 45\degree$ relative to the sample’s surface (and the optical axis). Each measurement set of all magnetic field magnitudes was repeated for different azimuthal angles $\phi=[-40\degree,80\degree]$. In addition, the magnetic field measurement order was randomized for each set independently to filter systematic background noises.

\begin{figure}[tbh]
    \includegraphics[width = 0.9 \linewidth]{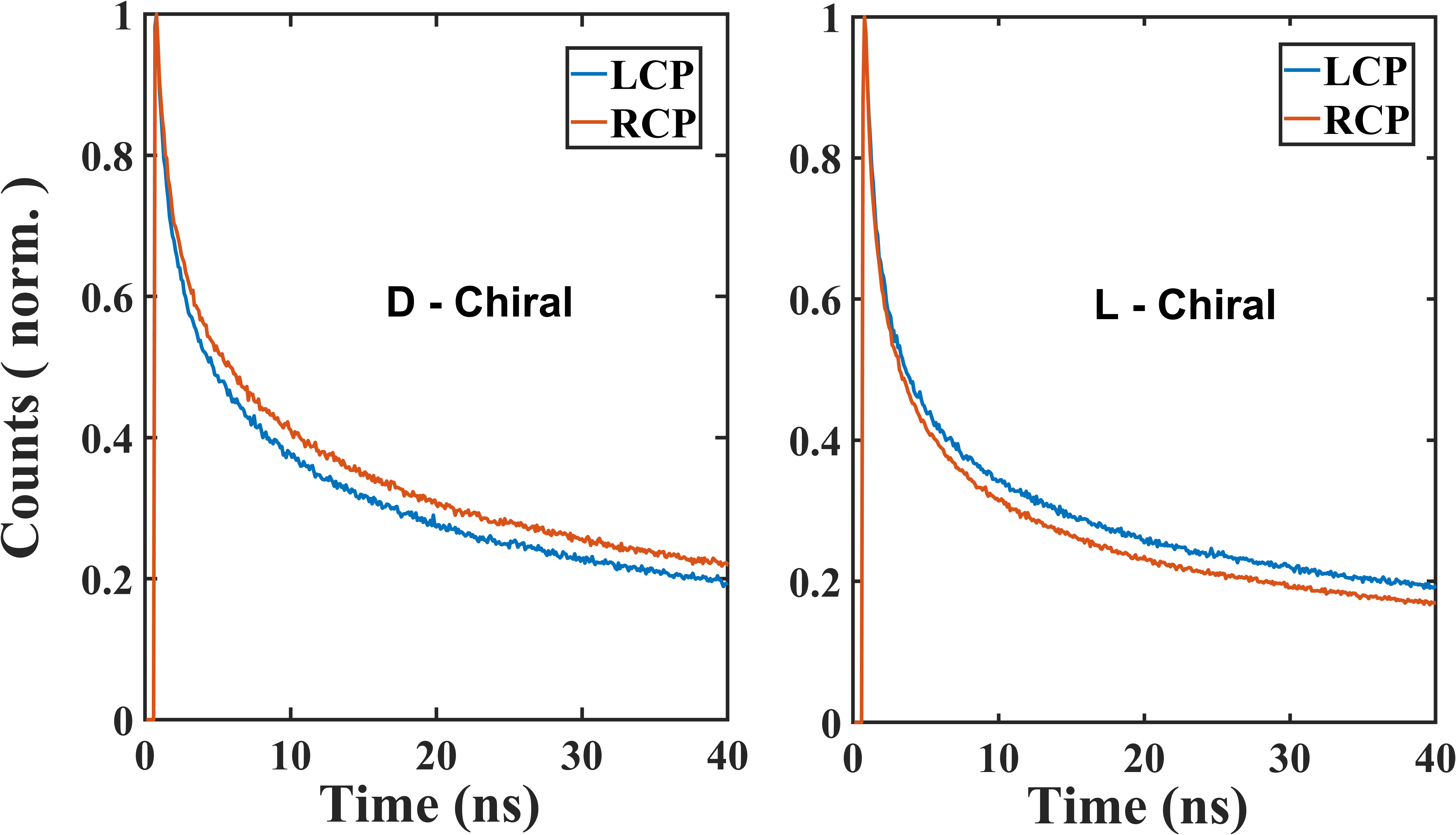}
    \caption{Symmetry breaking in a chiral system. Comparison of photoluminescence lifetime decay under a magnetic field of $B = 280$ Gauss for a multilayer of D-chiral or L-chiral QD assemblies, excited by LCP or RCP light.}
    \label{fig:PL_lifetime}
\end{figure}

Fig.~\ref{fig:PL_lifetime} shows the PL lifetime decay of the D-chiral and L-chiral samples using left circular polarization (LCP) and right circular polarization (RCP) light under a magnetic field of $B = 280$ Gauss, oriented at an azimuthal angle $\phi = 0\degree$. These results clearly show that the lifetime decay is longer when excited by RCP light for the D-chiral sample and shorter for the L-chiral system (and vice-versa). This indicates a CISS symmetry breaking in the chiral multilayer structure. 

The lifetime results were carefully analyzed and fitted to a biexponential decay model, taking into consideration the instrument response function (IRF) of the single photon detector. Data was extracted using the TCSPC reconvolution technique \cite{lakowicz2006principles}. The fitting function were thus:
\begin{align}
    PL = &A \exp{\left( \frac{-{t-t_0}}{\tau_1} \right) } \cdot \erfc{\left( \frac{-t-t_0- s^2/\tau_1}{\sqrt{2}s} \right)} +\\ \nonumber
    &B \exp{\left(\frac{-t-t_0}{\tau_2}\right)} \cdot \erfc{ \left( \frac{-t-t_0- s^2/\tau_2}{\sqrt{2}s} \right)} + C,
\end{align}
where $A, B$ are amplitudes and $\tau_1, \tau_2$ are the decay times for the long and short decays. The $\erfc(z)=1-\erf(z)$ is the complementary error function, with the IRF response width $s$ and the delay $t_0$ extracted from a calibration measurement. C is a background normalization offset dc constant.

\begin{figure}[tbh]
    \includegraphics[width = 0.9 \linewidth]{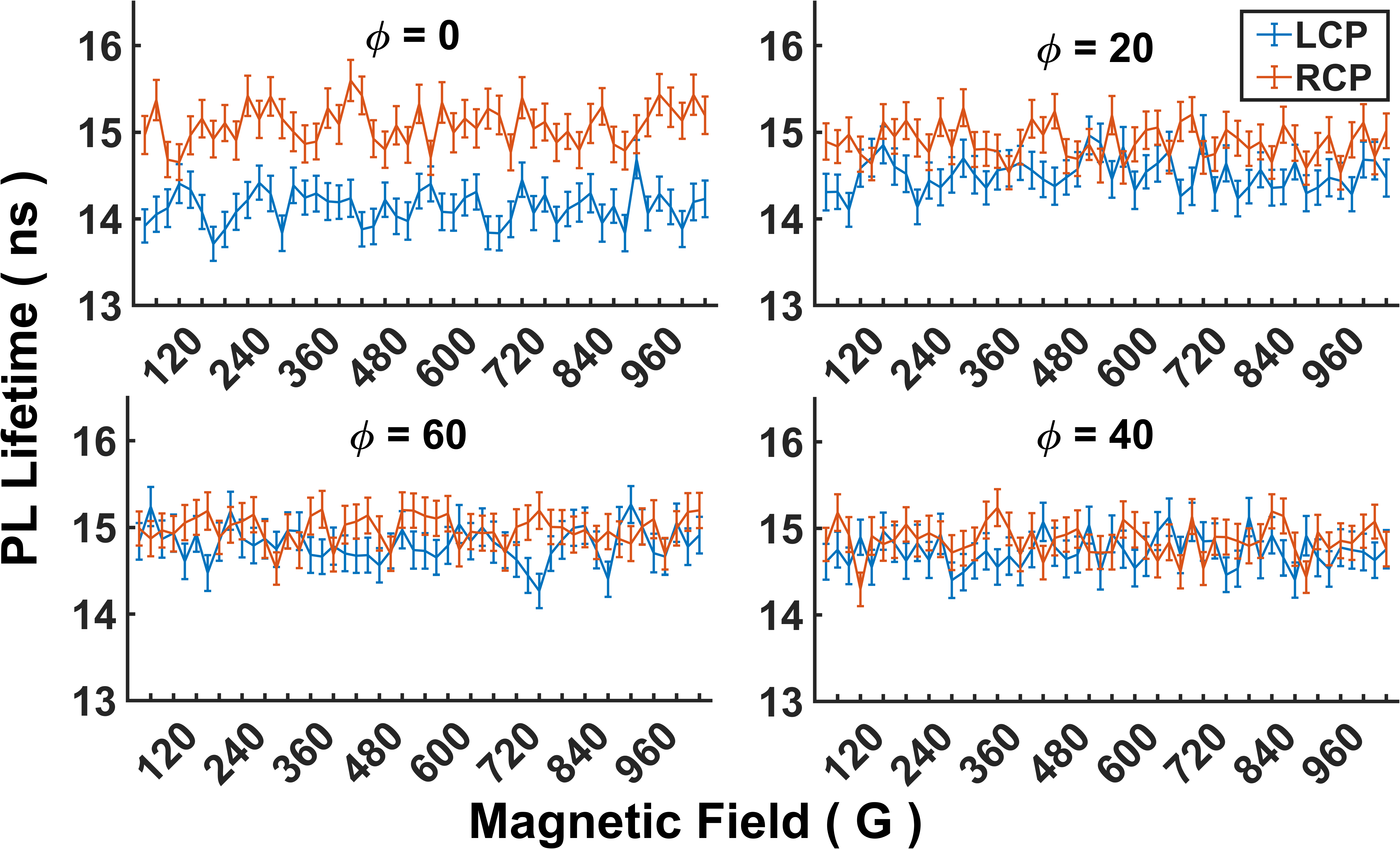}
    \caption{Lifetimes constants for the D-chiral sample. Lifetime constants extracted for LCP and RCP light are plotted in blue and red respectively, as a function of the applied magnetic field. The magnetic field is oriented in a polar angle $\theta = 45\degree$ relative to the sample surface. The magnet's azimuthal angles $\phi$ are aligned to (a) $\phi = 0\degree$; (b) $\phi = 20\degree$; (c) $\phi = 40\degree$ and (d) $\phi = 60\degree$  }
    \label{fig:phi_compare_D}
\end{figure}

This approach allows for the deconvolution of the complex decay profiles, separating between the fast and slow decay components. In Fig.~\ref{fig:phi_compare_D}, we present the long component lifetime constants extracted from the lifetime fitting results at different magnetic fields and azimuthal angles for the D-chiral QD multilayer sample. The results show that using RCP light, the PL lifetimes have longer decay rates compared to those when using LCP light. In addition, we notice an oscillating behavior of the PL lifetime as the magnetic field is changed, where the RCP and LCP lifetime dynamics are complementary to each other. Moreover, the difference between the lifetimes when using RCP and LCP light is highly dependent on the azimuthal angle $\phi$, strengthening the assumption that the alignment of the magnetic field relative to the axis of the chiral molecules is an important factor in the CISS dynamics.
A similar analysis (Supporting Information) was performed for the L-chiral sample, showing an inversion of the dynamics as expected. In particular, the lifetime constants for the L-chiral sample are longer for the LCP light. In addition, an oscillatory trend and an azimuthal angle dependence on $\phi$ are observed.

\begin{figure}[tbh]
    \includegraphics[width = 0.9 \linewidth]
{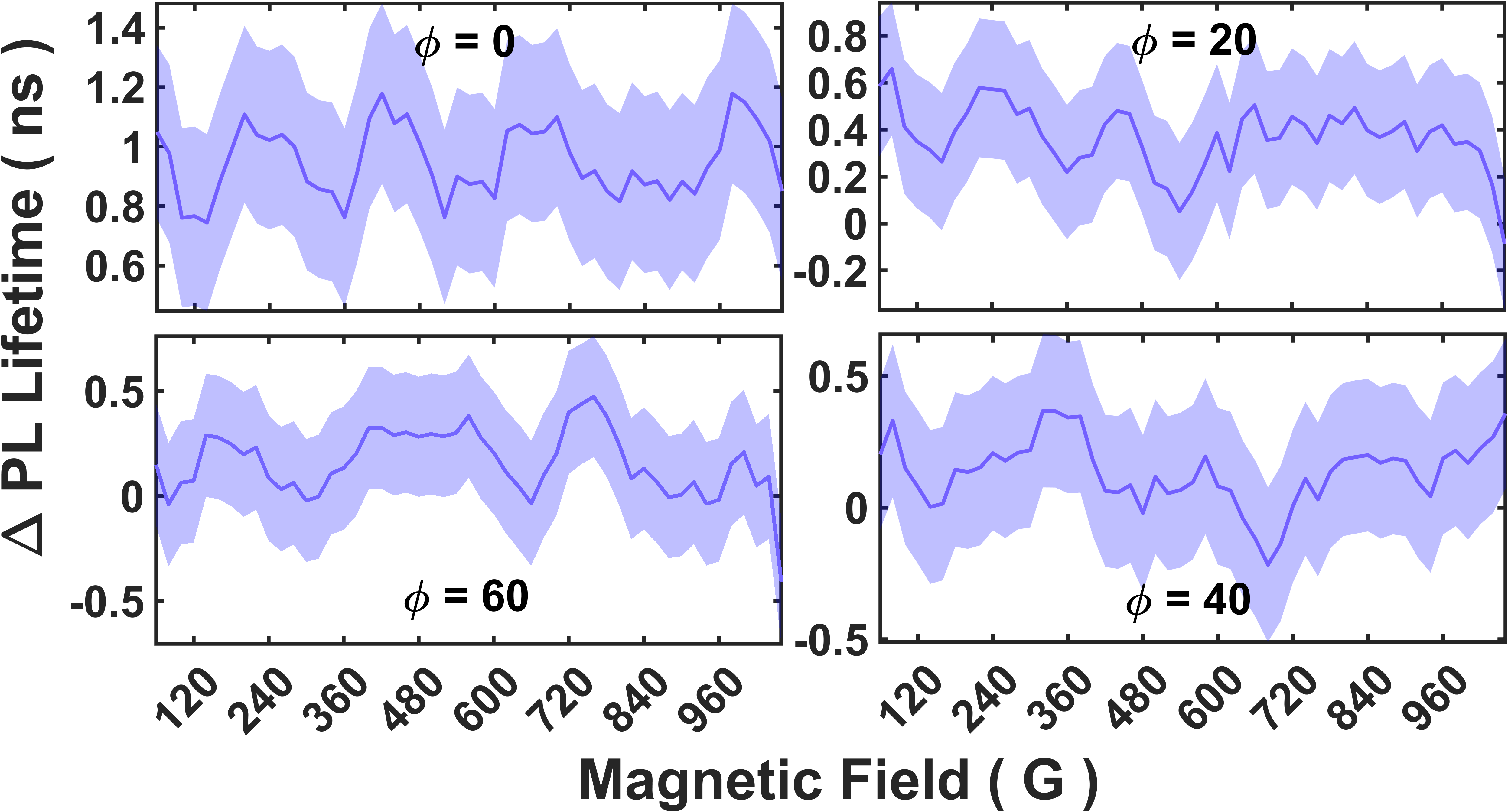}
    \caption{PL lifetime differences between RCP and LCP excitation light as a function of the magnetic field for the D-chiral sample. The magnetic field is oriented in a polar angle $\theta = 45\degree$ relative to the sample surface and the optical axis. The magnet is rotated around the sample in different azimuthal angles (a) $\phi = 0\degree$; (b) $\phi = 20\degree$; (c) $\phi = 40\degree$ and (d) $\phi = 60\degree$.}
    \label{fig:PL_diff_as_fun_of_B_D}
\end{figure}

The oscillatory behavior is highlighted by the difference of the lifetime decay rates between RCP and LCP excitation, defined as $\Delta PL_{\text{lifetime}} = PL_{\text{RCP}} - PL_{\text{LCP}}$. The results for the D-chiral sample are presented in Fig.  ~\ref{fig:PL_diff_as_fun_of_B_D}. The extracted lifetime differences exhibit an oscillatory behavior as a function of the magnetic field, consistent with the expected modulation arising from spin precession along the chiral axis. Moreover, the position of the first peak shifts as the azimuthal angle $\phi$ changes, further indicating the angular dependence of the transverse magnetic-field projection. 
A similar behavior is observed for the L-chiral sample (see Supporting Information Fig.~\ref*{fig:L_chiral_diff_angels}).

\begin{figure}[tbh]
   \includegraphics[width = 0.7 \linewidth]{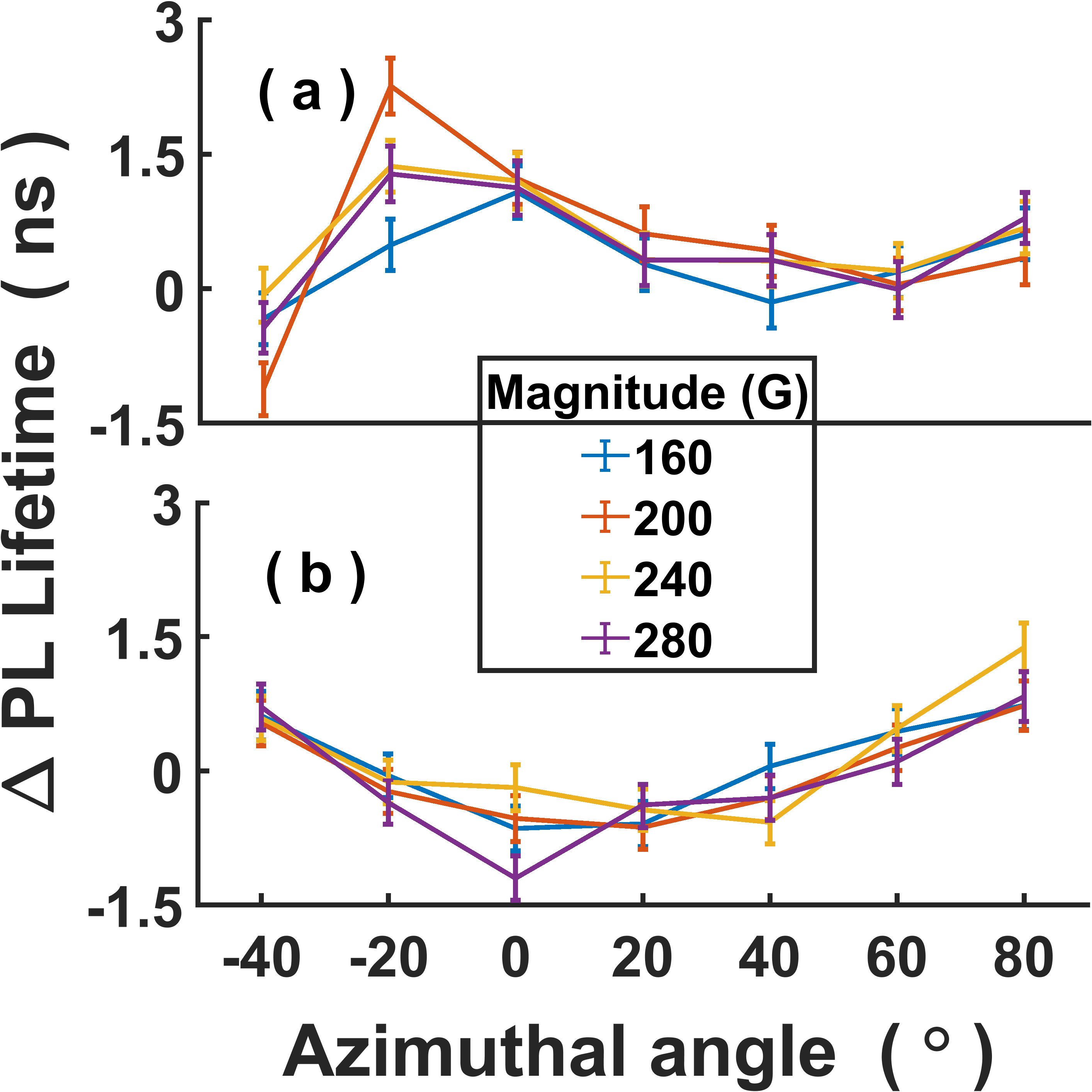}
\caption{PL lifetime differences between RCP and LCP light as a function of the azimuthal angle $\phi$, plotted for magnetic fields B = 160 (blue), 200 (red), 240 (yellow) and 280 (purple) Gauss. (a) D-chiral sample; (b) L-chiral sample.} 
\label{fig:PL_diff}
\end{figure}

In addition, we plot the lifetime difference $\Delta PL_{lifetime}$ for different azimuthal angles $\phi$, emphasizing the dependence of the lifetime decay rate on the magnetic field orientation. This is clearly shown in Fig.~\ref{fig:PL_diff}(a) and (b) for the D-chiral and L-chiral samples, respectively. Notably, the lifetime difference $\Delta PL_{\text{lifetime}}$ exhibits a periodic modulation with the azimuthal angle $\phi$, consistent with the expected geometric dependence of the transverse component of the magnetic field.

As illustrated in Fig.~\ref{fig:Illustration_and_QD_chiral_sample} (a), the spin precesses around the magnetic field direction. When the spin precession aligns with the chiral axis, $\Delta PL_{lifetime}$ varies with the spin oscillation. This effect arises from variations in spin dynamics due to the coupling between quantum dots and chiral molecules, enabling the detection of spin precession, which can be either enhanced or suppressed depending on the spin state. Specifically, maximum differences are expected at varying magnetic field strengths, determined by the projection of the field onto the axis perpendicular to the chiral system, exhibiting a $\cos(\phi)$ dependence\cite{das_insights_2024,moharana_chiral-induced_nodate}. Consequently, the lifetime difference changes with the azimuthal angle. As the angle varies, the difference between the lifetime decay times increases and decreases in a periodic manner. On this basis, the lifetime difference is expected to exhibit a characteristic angular modulation, consistent with the experimentally observed periodic behavior. An opposite trend is observed for the L-chiral linker (Fig.~\ref{fig:PL_diff}(b)).

\section{Discussion and model}
The spin selectivity effect in chiral structures is clearly demonstrated in Fig.~\ref{fig:PL_lifetime}, where selective excitation of a specific spin state using circularly polarized light induces a pronounced breaking of symmetry in the system. The effect was then explored in the presence of a magnetic field, revealing a magnetic field dependent oscillation of the PL decay times as presented in \ref{fig:phi_compare_D}.

When an external magnetic field is applied, the electron spin in each QD undergoes Larmor precession around the field direction. A critical parameter governing the resulting dynamics is the angle between the molecular (chiral) axis and the magnetic-field direction. As long as the transverse component of the magnetic field is significant,it drives the spin to rotate along the chiral axis. This coherent spin evolution periodically enhances and suppresses the preferred delocalization pathway, leading to a modulation of the effective PL lifetime. Previous studies have shown that in similar systems the chiral molecules align at a stable polar angle with respect to the sample surface \cite{peer_nanoscale_2015,meirzada_long-time-scale_2021}. Therefore, we choose to work with a magnetic field aligned at a fixed polar angle $\theta = 45\degree$.

By sweeping the azimuthal angle $\phi$, the perpendicular projection of the magnetic field relative to the chiral axis changes. Although spin precession occurs for any magnetic-field orientation, its influence on the PL lifetime is observable only through the component of the precession that projects onto the chiral axis. This projection is set by the transverse field component, whose magnitude determines the effective Larmor precession frequency along the chiral frame. 

Consequently, a larger perpendicular projection leads to a higher precession frequency and therefore to a faster (though experimentally averaged out) modulation of the PL lifetime.
This explains why changing the azimuthal angle $\phi$ modifies $\Delta PL_{\text{lifetime}}$, as shown in Fig.~\ref{fig:PL_diff_as_fun_of_B_D}. 
These observations indicate that the electron spin undergoes precession inside the QDs, and that only the component of this precession projecting onto the chiral molecular axis contributes to the measurable PL-lifetime modulation. Notably, both samples exhibit a periodic trend consistent with the expected $\cos(\phi)$ dependence, as shown in Fig.~\ref{fig:PL_diff}.

It is important to note that a clean oscillatory signature in the PL decay itself is not observed experimentally. This can be attributed to structural inhomogeneity in the multilayer film: not all QDs experience the same degree of chiral coupling and some QDs may not be chirally coupled at all due to size dispersion, symmetric configurations, or local structural disorder. As a result, the effective PL lifetime contains a distribution of decay constants, averaging out the oscillations and preventing the appearance of a well-resolved periodic signal in the raw PL decay curves. However, upon analyzing the extracted long-lived PL component under different magnetic field magnitudes and angles, the influence of spin precession becomes much clearer, as evidenced by the magnetic-field-dependent modulation of the lifetime. Despite this, the detailed origin of the observed oscillatory behavior is not straightforward and requires further theoretical modeling.

To gain deeper insight into these results and support our physical interpretation, we developed a model describing the coupled spin and decay dynamics in the chiral–QD system. The central feature of the model is the spin-selective coupling introduced by the chiral molecular linkers, which preferentially delocalizes one electronic spin orientation and therefore produces two distinct decay channels corresponding to lifetimes $\tau_{\uparrow}$ and $\tau_{\downarrow}$ of the different electron spins, respectively. To simulate the coupling inhomogeneities of the experimental system, the lifetimes are randomly generated from a normal distribution, using the measured lifetimes as the mean value, and choosing a standard deviation of $1$ ns. At any given moment, the probability of occupying the spin-up or spin-down state is determined by the instantaneous precession phase, $\varphi(t)$:
\[
P_{\uparrow}(t) = \cos^2 \left[\varphi(t)\right], \qquad  
P_{\downarrow}(t) = \sin^2 \left[\varphi(t)\right].
\]

The corresponding populations decay according to their respective lifetimes $\tau_{\uparrow}$ and $\tau_{\downarrow}$. For each applied magnetic field $B$, the precession phase evolves in time as
\[
\ \varphi(t) = \omega t + \varphi_{0},
\]
Where $\omega = \gamma B$  is the Larmor frequency determined by the magnitude of the magnetic field, with the proportionality constant $\gamma$ being the effective gyromagnetic ratio of the QD's electronic spin \cite{semina_electron_2020,hu_origin_2019} .
The initial spin state $\varphi_{0}$ is set in the experiment by the circular polarization of the excitation light, and is therefore either $0$ or $\pi/2$.

Starting from an initial population of $N$ particles initialized to either of the spin states, the number of decays at each time step is calculated using Poisson statistics, weighted by the time-dependent probabilities of occupying the spin-up or spin-down state. 
In this way, the model constructs the full PL decay under the influence of the magnetic field, allowing the long-lived lifetime component to be extracted in the same manner as in the experimental analysis. This results in a modulated PL decay that exhibits a reduced oscillatory behavior in the PL lifetime decay profile. Notably, this reduction in oscillations is very similar to the weak oscillatory signatures observed experimentally, as shown in the further analysis provided in the Supporting Information.

\begin{figure}[tbh]
    \includegraphics[width = 0.9 \linewidth]{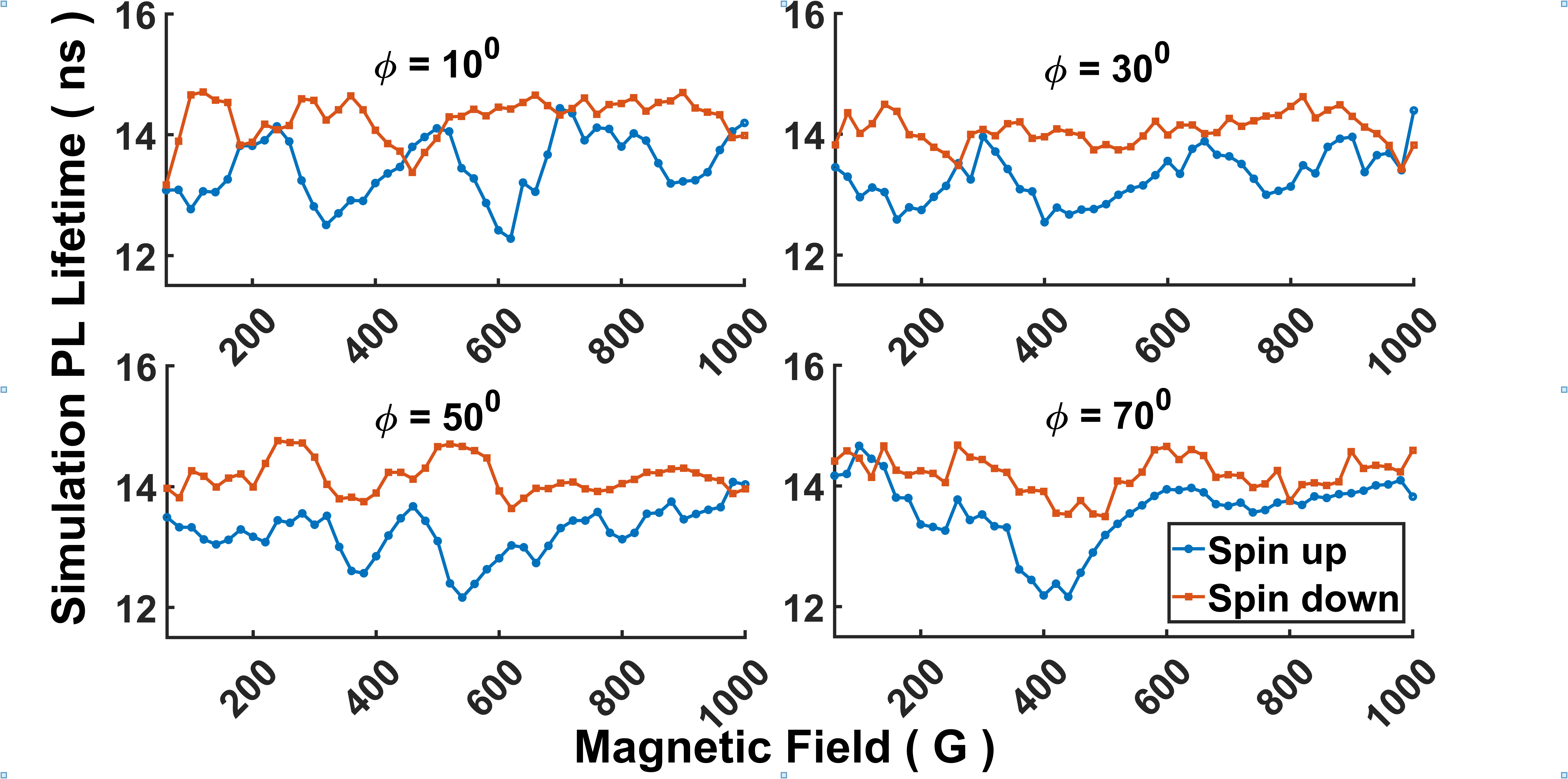}  
    \caption{Simulated long-lived PL lifetime decay rates for initially spin-up and spin-down states, corresponding to RCP and LCP excitation, respectively. The extracted lifetime decay rates are shown as a function of the applied magnetic field, with spin-up (blue) and spin-down (red) contributions. The magnetic field is applied at a polar angle of $\theta = 45\degree$ relative to the chiral molecular axis, and the azimuthal angle is set to (a) $\phi = 10\degree$, (b) $\phi = 30\degree$, (c) $\phi = 50\degree$, and (d) $\phi = 70\degree$.
    }
    \label{fig:Simulaion_of_D_chiral_combined_angle}
\end{figure}

In the model, the azimuthal angle $\phi$ is incorporated by calculating the angle $\alpha$ between the chiral molecular axis and the magnetic-field direction, such that the transverse field component driving spin precession is given by
\[
B_{\perp} = |\mathbf{B}|\sin\alpha = |\mathbf{B}|\sqrt{1-(\mathbf{n}\cdot\hat{\mathbf{B}})^2}.
\]
Using this framework, we simulate the PL decay for a range of magnetic-field values and different azimuthal angles, extracting the long-lived lifetime component $\tau$ using the same methods as for the experimental data. Comparing the experimental lifetimes to the simulated ones as a function of the magnetic field magnitude leads to a clear quantitative correspondence, highlighting the coherent dynamics of the electronic spin in the chiral-QD system. In addition, the dependence on the azimuthal angle is reflected in the modulation of the lifetime difference between LCP and RCP excitation. The variations in $\Delta PL_{\text{lifetime}}$ follow the expected angular dependence. This behavior is shown in Fig.~\ref{fig:Simulaion_of_D_chiral_combined_angle}, where simulated long-lived lifetimes for both initial spin states and different azimuthal angles are presented as a function of the magnetic field magnitude.

Overall, the simulations capture the key experimental trends observed in the azimuthal and magnetic field dependent PL lifetimes. In particular, the model reproduces the modulation of the PL lifetime difference, as well as the systematic reduction and re-enhancement of the modulation amplitude at larger azimuthal angles $\phi$. These effects arise naturally from the geometric projection of the transverse component of the magnetic field onto the chiral molecular axis, which governs the efficiency of spin precession during carrier relaxation.

While structural inhomogeneity and a distribution of chiral coupling strengths suppress clear oscillations in the raw PL decay, the extracted long-lived lifetime component remains sensitive to the underlying spin dynamics. The agreement between experiment and simulation therefore supports a physical picture in which spin-selective excitation, combined with chiral-mediated spin precession, govern the observed lifetime modulation. This behavior constitutes a clear signature of coherent spin-selective effects in the chiral quantum dot system.

\section{Summary}
In this work, symmetry breaking is demonstrated by assembling multilayer quantum dots (QDs) connected via chiral linkers. This configuration enables the observation of electron spin precession in the QDs, providing evidence for coherent spin-polarized electrons undergoing a spin-selective process that preserves their initial phase. 
The combined experimental and simulated analysis demonstrates that the observed lifetime modulation originates from the interplay between spin-selective excitation and coherent chiral-mediated spin precession in the presence of a transverse magnetic field. These results establish chiral QD assemblies as a room-temperature platform for probing coherent manifestations of the chiral-induced spin selectivity (CISS) effect, deepening our understanding of spin dynamics in chiral systems and highlighting their potential for future coherent quantum and spintronic applications.

\begin{acknowledgement}
The authors thank Guy Ron for fruitful discussions regarding the statistical analysis of the data.

N.B. acknowledges support by the European Commission’s Horizon Europe Framework Programme under the Research and Innovation Action GA No. 101070546-MUQUABIS and ERC CoG Project QMAG (no. 101087113). N.B. also acknowledges financial support by the Carl Zeiss Stiftung (HYMMS wildcard), the Ministry of Science and Technology, Israel, the innovation authority (project no. 70033), and the ISF (Grants No. 1380/21 and No. 3597/21).
Y.P. acknowledges the support of the NSF/BSF grant no. 2024621,  MAFAT grant no. 4441440446, and the Ministry of Innovation, Science and Technology, Israel grant no. 1001578129.  H.T.F. acknowledges support from the Ministry of Science and Technology of Israel through the Aloni PhD Fellowship.
\end{acknowledgement}

\begin{suppinfo}
The Supporting Information is available free of charge at \ldots

\begin{itemize}
  \item Experimental setup, additional measurements, quantum dots characterization, further statistical analysis.
\end{itemize}

\end{suppinfo}

\bibliography{references/references_for_lifetime_paper}

@article{das_insights_2024,
	title = {Insights into the {Mechanism} of {Chiral}-{Induced} {Spin} {Selectivity}: {The} {Effect} of {Magnetic} {Field} {Direction} and {Temperature}},
	volume = {36},
	issn = {0935-9648},
	url = {https://doi.org/10.1002/adma.202313708},
	doi = {10.1002/adma.202313708},
	number = {29},
	urldate = {2026-01-05},
	journal = {Advanced Materials},
	author = {Das, Tapan Kumar and Naaman, Ron and Fransson, Jonas},
	month = jul,
	year = {2024},
	note = {Publisher: John Wiley \& Sons, Ltd},
	pages = {2313708},
}

@article{mollers_chirality_2022,
	title = {Chirality {Induced} {Spin} {Selectivity} – the {Photoelectron} {View}},
	volume = {62},
	issn = {0021-2148},
	url = {https://doi.org/10.1002/ijch.202200062},
	doi = {10.1002/ijch.202200062},
	number = {11-12},
	urldate = {2026-01-05},
	journal = {Israel Journal of Chemistry},
	author = {Möllers, Paul V. and Göhler, Benjamin and Zacharias, Helmut},
	month = dec,
	year = {2022},
	note = {Publisher: John Wiley \& Sons, Ltd},
	pages = {e202200062},
}

@article{hu_origin_2019,
	title = {Origin of {Two} {Larmor} {Frequencies} in the {Coherent} {Spin} {Dynamics} of {Colloidal} {CdSe} {Quantum} {Dots} {Revealed} by {Controlled} {Charging}},
	volume = {10},
	url = {https://doi.org/10.1021/acs.jpclett.9b01534},
	doi = {10.1021/acs.jpclett.9b01534},
	number = {13},
	journal = {The Journal of Physical Chemistry Letters},
	author = {Hu, Rongrong and Yakovlev, Dmitri R. and Liang, Pan and Qiang, Gang and Chen, Cong and Jia, Tianqing and Sun, Zhenrong and Bayer, Manfred and Feng, Donghai},
	month = jul,
	year = {2019},
	note = {Publisher: American Chemical Society},
	pages = {3681--3687},
}

@article{semina_electron_2020,
	title = {Electron, hole and exciton effective g-factors in semiconductor nanocrystals},
	journal = {arXiv},
	author = {Semina, Mikhail A. and Golovatenko, Alexander A. and Rodina, Alexey V.},
	year = {2020},
	note = {\_eprint: 2011.11041},
}

@article{aiello_chirality-based_2022,
	title = {A {Chirality}-{Based} {Quantum} {Leap}},
	volume = {16},
	issn = {1936-0851},
	url = {https://doi.org/10.1021/acsnano.1c01347},
	doi = {10.1021/acsnano.1c01347},
	number = {4},
	journal = {ACS Nano},
	author = {Aiello, Clarice D. and Abendroth, John M. and Abbas, Muneer and Afanasev, Andrei and Agarwal, Shivang and Banerjee, Amartya S. and Beratan, David N. and Belling, Jason N. and Berche, Bertrand and Botana, Antia and Caram, Justin R. and Celardo, Giuseppe Luca and Cuniberti, Gianaurelio and Garcia-Etxarri, Aitzol and Dianat, Arezoo and Diez-Perez, Ismael and Guo, Yuqi and Gutierrez, Rafael and Herrmann, Carmen and Hihath, Joshua and Kale, Suneet and Kurian, Philip and Lai, Ying-Cheng and Liu, Tianhan and Lopez, Alexander and Medina, Ernesto and Mujica, Vladimiro and Naaman, Ron and Noormandipour, Mohammadreza and Palma, Julio L. and Paltiel, Yossi and Petuskey, William and Ribeiro-Silva, João Carlos and Saenz, Juan José and Santos, Elton J. G. and Solyanik-Gorgone, Maria and Sorger, Volker J. and Stemer, Dominik M. and Ugalde, Jesus M. and Valdes-Curiel, Ana and Varela, Solmar and Waldeck, David H. and Wasielewski, Michael R. and Weiss, Paul S. and Zacharias, Helmut and Wang, Qing Hua},
	month = apr,
	year = {2022},
	note = {Publisher: American Chemical Society},
	pages = {4989--5035},
}

@article{pirro_advances_2021,
	title = {Advances in coherent magnonics},
	volume = {6},
	issn = {2058-8437},
	url = {https://doi.org/10.1038/s41578-021-00332-w},
	doi = {10.1038/s41578-021-00332-w},
	number = {12},
	journal = {Nature Reviews Materials},
	author = {Pirro, Philipp and Vasyuchka, Vitaliy I. and Serga, Alexander A. and Hillebrands, Burkard},
	month = dec,
	year = {2021},
	pages = {1114--1135},
}

@article{meirzada_long-time-scale_2021,
	title = {Long-{Time}-{Scale} {Magnetization} {Ordering} {Induced} by an {Adsorbed} {Chiral} {Monolayer} on {Ferromagnets}},
	volume = {15},
	issn = {1936-0851},
	url = {https://doi.org/10.1021/acsnano.1c00455},
	doi = {10.1021/acsnano.1c00455},
	number = {3},
	journal = {ACS Nano},
	author = {Meirzada, I. and Sukenik, N. and Haim, G. and Yochelis, S. and Baczewski, L. T. and Paltiel, Y. and Bar-Gill, N.},
	month = mar,
	year = {2021},
	note = {Publisher: American Chemical Society},
	pages = {5574--5579},
}

@book{lakowicz2006principles,
	edition = {3},
	title = {Principles of fluorescence spectroscopy},
	publisher = {Springer},
	author = {Lakowicz, Joseph R.},
	year = {2006},
	doi = {10.1007/978-0-387-46312-4},
}

@article{becker2005tcspc,
	title = {Advanced time-correlated single photon counting techniques},
	volume = {247},
	doi = {10.1111/j.1365-2818.2011.03537.x},
	journal = {Journal of Microscopy},
	author = {Becker, Wolfgang},
	year = {2012},
	pages = {119--136},
}

@article{naaman2020acc,
	title = {Chiral induced spin selectivity gives a new twist on spin-control in chemistry},
	volume = {53},
	doi = {10.1021/acs.accounts.0c00485},
	number = {11},
	journal = {Accounts of Chemical Research},
	author = {Naaman, Ron and Paltiel, Yossi and Waldeck, David H.},
	year = {2020},
	pages = {2659--2667},
}

@article{decher1997layer,
	title = {Fuzzy nanoassemblies: {Toward} layered polymeric multicomposites},
	volume = {277},
	doi = {10.1126/science.277.5330.1232},
	journal = {Science},
	author = {Decher, Gero},
	year = {1997},
	pages = {1232--1237},
}

@article{warburton2013single,
	title = {Single spins in self-assembled quantum dots},
	volume = {12},
	doi = {10.1038/nmat3462},
	journal = {Nature Materials},
	author = {Warburton, Richard J.},
	year = {2013},
	pages = {483--493},
}

@article{lodahl2015interfacing,
	title = {Interfacing single photons and single quantum dots with photonic nanostructures},
	volume = {87},
	doi = {10.1103/RevModPhys.87.347},
	journal = {Reviews of Modern Physics},
	author = {Lodahl, Peter and Mahmoodian, Sahand and Stobbe, Soren},
	year = {2015},
	pages = {347--400},
}

@article{zwanenburg2013silicon,
	title = {Silicon quantum electronics},
	volume = {85},
	doi = {10.1103/RevModPhys.85.961},
	journal = {Reviews of Modern Physics},
	author = {Zwanenburg, Floris A. and {others}},
	year = {2013},
	pages = {961--1019},
}

@article{press2008complete,
	title = {Complete quantum control of a single quantum dot spin using ultrafast optical pulses},
	volume = {456},
	doi = {10.1038/nature07530},
	journal = {Nature},
	author = {Press, David and Ladd, Thaddeus D. and Zhang, Bingyang and Yamamoto, Yoshihisa},
	year = {2008},
	pages = {218--221},
}

@book{michler2009single,
	title = {Single semiconductor quantum dots},
	publisher = {Springer},
	author = {Michler, Peter},
	year = {2009},
	doi = {10.1007/978-3-540-77846-6},
}

@article{loss1998quantum,
	title = {Quantum computation with quantum dots},
	volume = {57},
	doi = {10.1103/PhysRevA.57.120},
	journal = {Physical Review A},
	author = {Loss, Daniel and DiVincenzo, David P.},
	year = {1998},
	pages = {120--126},
}

@article{xie_spin_2011,
	title = {Spin {Specific} {Electron} {Conduction} through {DNA} {Oligomers}},
	volume = {11},
	issn = {1530-6984},
	url = {https://doi.org/10.1021/nl2021637},
	doi = {10.1021/nl2021637},
	number = {11},
	journal = {Nano Letters},
	author = {Xie, Zouti and Markus, Tal Z. and Cohen, Sidney R. and Vager, Zeev and Gutierrez, Rafael and Naaman, Ron},
	month = nov,
	year = {2011},
	note = {Publisher: American Chemical Society},
	pages = {4652--4655},
}

@article{https://doi.org/10.1002/smll.201804557,
	title = {Single domain 10 nm ferromagnetism imprinted on superparamagnetic nanoparticles using chiral molecules},
	volume = {15},
	url = {https://onlinelibrary.wiley.com/doi/abs/10.1002/smll.201804557},
	doi = {https://doi.org/10.1002/smll.201804557},
	number = {1},
	journal = {Small},
	author = {Koplovitz, Guy and Leitus, Gregory and Ghosh, Supriya and Bloom, Brian P. and Yochelis, Shira and Rotem, Dvir and Vischio, Fabio and Striccoli, Marinella and Fanizza, Elisabetta and Naaman, Ron and Waldeck, David H. and Porath, Danny and Paltiel, Yossi},
	year = {2019},
	note = {tex.eprint: https://onlinelibrary.wiley.com/doi/pdf/10.1002/smll.201804557},
	pages = {1804557},
}

@article{bloom_chiral_2024,
	title = {Chiral {Induced} {Spin} {Selectivity}},
	volume = {124},
	issn = {0009-2665},
	url = {https://doi.org/10.1021/acs.chemrev.3c00661},
	doi = {10.1021/acs.chemrev.3c00661},
	number = {4},
	journal = {Chemical Reviews},
	author = {Bloom, Brian P. and Paltiel, Yossi and Naaman, Ron and Waldeck, David H.},
	month = feb,
	year = {2024},
	note = {Publisher: American Chemical Society},
	pages = {1950--1991},
}

@article{naaman_chiral_2020,
	title = {Chiral {Induced} {Spin} {Selectivity} {Gives} a {New} {Twist} on {Spin}-{Control} in {Chemistry}},
	volume = {53},
	issn = {0001-4842},
	url = {https://doi.org/10.1021/acs.accounts.0c00485},
	doi = {10.1021/acs.accounts.0c00485},
	number = {11},
	journal = {Accounts of Chemical Research},
	author = {Naaman, Ron and Paltiel, Yossi and Waldeck, David H.},
	month = nov,
	year = {2020},
	note = {Publisher: American Chemical Society},
	pages = {2659--2667},
}

@article{annurev:/content/journals/10.1146/annurev-physchem-040214-121554,
	title = {Spintronics and chirality: {Spin} selectivity in electron transport through chiral molecules},
	volume = {66},
	issn = {1545-1593},
	url = {https://www.annualreviews.org/content/journals/10.1146/annurev-physchem-040214-121554},
	doi = {https://doi.org/10.1146/annurev-physchem-040214-121554},
	number = {Volume 66, 2015},
	journal = {Annual Review of Physical Chemistry},
	author = {Naaman, Ron and Waldeck, David H.},
	year = {2015},
	note = {Publisher: Annual Reviews
Type: Journal Article},
	pages = {263--281},
}

@article{liu_coherent_2019,
	title = {Coherent quantum control of nitrogen-vacancy center spins near 1000 kelvin},
	volume = {10},
	issn = {2041-1723},
	url = {https://doi.org/10.1038/s41467-019-09327-2},
	doi = {10.1038/s41467-019-09327-2},
	number = {1},
	journal = {Nature Communications},
	author = {Liu, Gang-Qin and Feng, Xi and Wang, Ning and Li, Quan and Liu, Ren-Bao},
	month = mar,
	year = {2019},
	pages = {1344},
}

@article{lin_room-temperature_2023,
	title = {Room-temperature coherent optical manipulation of hole spins in solution-grown perovskite quantum dots},
	volume = {18},
	issn = {1748-3395},
	url = {https://doi.org/10.1038/s41565-022-01279-x},
	doi = {10.1038/s41565-022-01279-x},
	number = {2},
	journal = {Nature Nanotechnology},
	author = {Lin, Xuyang and Han, Yaoyao and Zhu, Jingyi and Wu, Kaifeng},
	month = feb,
	year = {2023},
	pages = {124--130},
}

@article{doi:10.1126/science.ads0512,
	title = {Ultrafast all-optical coherence of molecular electron spins in room-temperature water solution},
	volume = {386},
	url = {https://www.science.org/doi/abs/10.1126/science.ads0512},
	doi = {10.1126/science.ads0512},
	number = {6724},
	journal = {Science},
	author = {Sutcliffe, Erica and Kazmierczak, Nathanael P. and Hadt, Ryan G.},
	year = {2024},
	note = {tex.eprint: https://www.science.org/doi/pdf/10.1126/science.ads0512},
	pages = {888--892},
}

@article{al-bustami_atomic_2022,
	title = {Atomic and {Molecular} {Layer} {Deposition} of {Chiral} {Thin} {Films} {Showing} up to 99\% {Spin} {Selective} {Transport}},
	volume = {22},
	issn = {1530-6984},
	url = {https://doi.org/10.1021/acs.nanolett.2c01953},
	doi = {10.1021/acs.nanolett.2c01953},
	number = {12},
	journal = {Nano Letters},
	author = {Al-Bustami, H. and Khaldi, S. and Shoseyov, O. and Yochelis, S. and Killi, K. and Berg, I. and Gross, E. and Paltiel, Y. and Yerushalmi, R.},
	month = jun,
	year = {2022},
	note = {Publisher: American Chemical Society},
	pages = {5022--5028},
}

@article{fridman_ultrafast_2023,
	title = {Ultrafast {Coherent} {Delocalization} {Revealed} in {Multilayer} {QDs} under a {Chiral} {Potential}},
	volume = {14},
	doi = {10.1021/acs.jpclett.2c03743},
	journal = {The journal of physical chemistry letters},
	author = {Fridman, Hanna and Levy, Hadar and Meir, Amitai and Casotto, Andrea and Malkinson, Rotem and Dehnel, Joanna and Yochelis, Shira and Lifshitz, Efrat and Collini, Elisabetta},
	month = feb,
	year = {2023},
	pages = {2234--2240},
}

@article{moharana_chiral-induced_nodate,
	title = {Chiral-induced unidirectional spin-to-charge conversion},
	volume = {11},
	url = {https://doi.org/10.1126/sciadv.ado4285},
	doi = {10.1126/sciadv.ado4285},
	number = {1},
	urldate = {2025-04-21},
	journal = {Science Advances},
	author = {Moharana, Ashish and Kapon, Yael and Kammerbauer, Fabian and Anthofer, David and Yochelis, Shira and Shema, Hadar and Gross, Elad and Kläui, Mathias and Paltiel, Yossi and Wittmann, Angela},
	note = {Publisher: American Association for the Advancement of Science},
	pages = {eado4285},
}

@article{peer_nanoscale_2015,
	title = {Nanoscale {Charge} {Separation} {Using} {Chiral} {Molecules}},
	volume = {2},
	url = {https://doi.org/10.1021/acsphotonics.5b00343},
	doi = {10.1021/acsphotonics.5b00343},
	number = {10},
	journal = {ACS Photonics},
	author = {Peer, Nir and Dujovne, Irene and Yochelis, Shira and Paltiel, Yossi},
	month = oct,
	year = {2015},
	note = {Publisher: American Chemical Society},
	pages = {1476--1481},
}

@article{fridman_spin-exciton_2019,
	title = {Spin-{Exciton} {Delocalization} {Enhancement} in {Multilayer} {Chiral} {Linker}/{Quantum} {Dot} {Structures}},
	volume = {10},
	url = {https://doi.org/10.1021/acs.jpclett.9b01433},
	doi = {10.1021/acs.jpclett.9b01433},
	number = {14},
	journal = {The Journal of Physical Chemistry Letters},
	author = {Fridman, Hanna T. and Dehnel, Johanna and Yochelis, Shira and Lifshitz, Efrat and Paltiel, Yossi},
	month = jul,
	year = {2019},
	note = {Publisher: American Chemical Society},
	pages = {3858--3862},
}

@article{bloom_chirality_2017,
	title = {Chirality {Control} of {Electron} {Transfer} in {Quantum} {Dot} {Assemblies}},
	volume = {139},
	issn = {0002-7863},
	url = {https://doi.org/10.1021/jacs.7b04639},
	doi = {10.1021/jacs.7b04639},
	number = {26},
	journal = {Journal of the American Chemical Society},
	author = {Bloom, Brian P. and Graff, Brittney M. and Ghosh, Supriya and Beratan, David N. and Waldeck, David H.},
	month = jul,
	year = {2017},
	note = {Publisher: American Chemical Society},
	pages = {9038--9043},
}

@article{bezen_chiral_2018,
	title = {Chiral {Molecule}-{Enhanced} {Extinction} {Ratios} of {Quantum} {Dots} {Coupled} to {Random} {Plasmonic} {Structures}},
	volume = {34},
	issn = {0743-7463},
	url = {https://doi.org/10.1021/acs.langmuir.8b00155},
	doi = {10.1021/acs.langmuir.8b00155},
	number = {9},
	journal = {Langmuir},
	author = {Bezen, Lior and Yochelis, Shira and Jayarathna, Dilhara and Bhunia, Dinesh and Achim, Catalina and Paltiel, Yossi},
	month = mar,
	year = {2018},
	note = {Publisher: American Chemical Society},
	pages = {3076--3081},
}

@book{klimov_victor_i_nanocrystal_nodate,
	edition = {Second edition},
	title = {Nanocrystal {Quantum} {Dots}},
	author = {Klimov, Victor I.},
}

@article{gohler_b_spin_2011,
	title = {Spin {Selectivity} in {Electron} {Transmission} {Through} {Self}-{Assembled} {Monolayers} of {Double}-{Stranded} {DNA}},
	volume = {331},
	url = {https://doi.org/10.1126/science.1199339},
	doi = {10.1126/science.1199339},
	number = {6019},
	urldate = {2021-11-08},
	journal = {Science},
	author = {{Göhler B.} and {Hamelbeck V.} and {Markus T. Z.} and {Kettner M.} and {Hanne G. F.} and {Vager Z.} and {Naaman R.} and {Zacharias H.}},
	month = feb,
	year = {2011},
	note = {Publisher: American Association for the Advancement of Science},
	pages = {894--897},
}

\end{document}